# The potential and challenges of Big data - Recommendation systems next level application

Fatima EL Jamiy[1], Abderrahmane Daif[2], Mohamed Azouazi[3] and Abdelaziz Marzak[4]

[1] Hassan II University, Faculty Of Sciences Ben m'Sik, Laboratoire Mathématiques Informatique et Traitement de l'Information MITI, Casablanca, Morocco

**Abstract**

The continuous increase of data generated provides enormous possibilities of both public and private companies. The management of this mass of data or big data will play a crucial role in the society of the future, as it finds applications in different fields. There are so much potential and extremely useful insights hidden in the huge volume of data. The advanced analysis techniques available including predictive analytics, text mining, semantic analysis are needed to enable organizations to create a competitive advantage through data analyzed with different levels of sophistication, speed and accuracy previously unavailable. Therefore, is it still possible to have that level of sophistication with the ubiquitous numeric ocean that accompanies use every day via connected devices that invade our lives?

However, development of big data requires a good understanding of the issues associated with it. And this is the purpose of this paper, which focuses on giving a close-up view of big data analysis, opportunities and challenges.

## 1. Introduction

Data befall exponentially on organizations. The new systems of information, expression vectors, payment systems, and opening multiple databases public and private data generated daily a stream of billions information. In addition, each actor (individual, administration, organization, business, community group) is producer of new corpus of non-or semi-structured information: personal data, geo-localized applications (together with the time dimension), conversations on social networks , events, dematerialized content, Flickr photos, microblogs ... On the other hand, submissions from the multiple digital communication devices (internet of Things, M2M communications, sensors, probes, RFID) produce the data on a large scale.

Big Data is the ability to integrate, synchronize, manage and valuate this data deluge, extremely with a great diversity of type. It is about taking advantage of massive data and to extract the essence and value. The data are no longer called structured and relational formats, but referenced as unstructured and heterogeneous content (comments, micro-focus, video, images, audio, sensor data. ...). The nature of these heterogeneous data has no limits for Big Data.

The way we take up in doing business, management and research has been changed with Big data. This huge amount of data has overcome our capability to process by traditional processing approaches that begin to show their limits. In fact, available data analysis tools cannot cope with the increase in size, variety and frequency of change of the Web. Therefore essential evolutions in the architecture of data-management systems are needed.

Taking into consideration the field of costumer management, in which an economy increasingly competitive, corporate wealth lies in their customers. The share of customer replaced the market share. Several studies show that the profitability of existing customers is much higher than new customers. It is cheaper to sell to existing customers than new customers.

The implementation of a retention strategy is based on a closer relationship with the customer. Managing customer relationships strategically uses information, processes, technology and people. We will focus on using technology to analyze information to learn more about the customer and improve sales. However, this approach to knowledge-based customer data analysis is hampered by the volume and variety of data that swell dramatically.

The paper is organized as follows: Section 2 focuses on the context and the emergence of Big Data; its opportunities and challenges; Section 3 is dedicated to the costumer management; usages and limits; the final Section draws the conclusion and future work.

## 2. Big data: opportunities and challenges

2.1 Emergence of Big Data

The constantly creation of diversified amount of data (pictures, videos, audio...) every day have led to the emergence of the concept of big data. The quick evolution of information technologies has strengthened the increase of volume Data. It has achieved a high quantity in every sector and function of the economy [3].

According to the IDC study [2] (Digital Universe study of International Data Corporation), the amount of information the world generated and replicated in 2009 was 800 exabytes, and surpassed 1.8 Zettabytes (ZB) in 2011, making an increase by a factor of 9 in just five years. They estimated an expansion rate of 40 percent to 2020.

The trend for tablets (63.6 million tablets sold worldwide in 2011), the explosion of Smartphone sales (nearly 10 million French fitted in 2010) and the proliferation of mobile access via Wi terminals Fi (in hotels, restaurants, shops ...) are the most representative examples [7].
Smart home, which includes the Internet connection of a multitude of household appliances (TV, refrigerator, cooktop, smart electric meter ...). All these connected products generate considerable amounts of data in their operations (trace calls, browsing history ...).

The generalization of internet use and the rise of new usages are identified as the principal origins of data deluge. Capturing almost every aspect of life is now possible. The world count today approximately 5 billion mobile phones and millions of sensors. Taking social networking as example, Facebook count 2,7 billion of « likes » or « I like » every day and 2,5 billion of contents (statuts, photos, videos) are shared every day» the social networking should store approximately 500 terabytes every day. In addition, 50 billion of tweets per day pass through Tweeter servers [5].

The growth of the number [6] of people that use internet explain also the explosion of data. Internet usage is increasing over the world, with 3.4 billion of users estimated in 2016, approximately 45 % of the worldwide population.
Web search engines (Google, Amazon, and Yahoo) are the first to face the problem of big volume of data to handle in real time. Therefore, they are the first to develop big data management tools and make them available to open sources communities [4].
In fact, there are many other factors participating every day to this proliferation:

- Technological advances,
- Social platforms,
- Multimedia platforms,
- Predominance of systems ERP and CRM.

2.2 The 5 Vs in Big Data analysis:

Gartner [8] uses the "3 Vs" to describe big data:

**Volume**: covers the size of the data needed for management. There is more data than ever before, its size keep on growing exponentially: 90% of all the data available today were created in the last two years [7]. A short time ago, we were talking about gigabytes, we are talking now relatively about terabytes, petabytes, exabytes and even zettabytes.
**Velocity**: describes the speed with which the data is generated and processed. We are focused on getting knowledge from the data arriving as streams in real time. More we focus on real time; more we are in big data problem. Gradually, the immediate treatment of data would be the key element of a model big data.
**Variety**: refers to the difference in the type of data we have collected. The data analyzed is not anymore structured as the anterior data, but could be text, pictures, multimedia content, digital traces, sensor data, etc. It is about the ability to give an additional value to the internal traditional data by combining it with a big variety of other external sources of data.
Nowadays, there are two more criteria:
**Veracity**: this is the most crucial point. Confidence in the accuracy of data collected and presented is an issue more and more real as the number of sources generating data increases. It reflects the emphasis of the need for quality data in a Big Data system.
**Value**: represents the value that can be drawn from these data and the uses they produce. Sorting data is then essential. It is essential to properly select the data to be analyzed, based on its activity and especially its objectives.

2.3 Big Data: potential and applications

The Big Data aim is to improve decisions and competitiveness for companies and the public administrations, which will create a significant growth of the world economy.
According to the report from McKinsey institute [1], Big data help to better listen to costumers, better understand their ways of using services and hence the offer. These applications also have benefits in many fields: from scientific researches to national security, and from global economy to society administration.

There are several applications of the Big Data problems, we present briefly the most widespread:

**Business and Commerce**: estimates have shown that in the era of information, every 1.2 years, the volume of business data worldwide doubles, across almost companies.
The insights hidden in this deluge of data can be discovered by using Big data, which help for example to optimize business processes, detect a pattern, better understand customers, anticipate their behaviors, needs and intentions [8].

**Science and Research:** as various types of data are generated and produced in the field of Science and research. These scientific domains are based on data-intensive scientific discovery. Taking CERN [1] as example, the Swiss nuclear physics lab with its Large Hadron Collider, the world's largest and most powerful particle accelerator. The computing powers of thousands of computers distributed across 150 data centers worldwide are used to process all the data, it produces.

**Health:** public health and medicine are fields in which employing effectively the healthcare data deluge would give the right outcomes to the patient and reduce care cost. The computing power of big data allow us to mine entire DNA strings in minutes and will provide us the possibility to discover, monitor, improve health aspects of every one and predict disease patterns [11].

**Smart cities:** as described in the work [9], cities and countries are currently being transformed by the new possibilities big data brings. Sustainable economic development and high quality of life, with wise management of natural resources are the major objectives of cities. For example, in many cities, the transport infrastructure and utility processes are all joined up in order to turn into smart cities. This will help them to minimize jams and optimize traffic flows.

2.4 Challenges

Opportunities contain challenges and when we talk about big data, its challenges are significant. Starting with the simple question about storage, capture and analysis, it gives the possibility to take advantage from the knowledge hidden behind these new information sources. Overcoming all those challenges is a big deal. Precisely, when we are facing a huge amount of data generated with a great speed that exceed our ability to exploit. We seek tools and platforms that combine and focus on the three dimensions of big data (volume, real-time, variety).
Up to now, and according to the recent related researches [3,8,10,11], adequate tools and platforms to harness totally the data deluge do not yet exist. In numerous applications big data, and more specifically real time analysis we are not able to resolve the real problem.
Data integration and volume**,** scalability, timeliness and data security are the main challenges in big data analysis. Making an effective representation, management, and processing of unstructured and semi-structured data remain a difficult challenge, more specifically when we take into account the variety of this data deluge.
Different challenges require to be addressed in order to build an effective solution big data. The most important are:

**Integration and acquisition:** every day we create 2.5 quintillion bytes of data, and this amount keeps rising exponentially. Just because of the lack of space to stock the data, different domain such as medicine and finance delete the data they create and collect.
The skills we use to collect and store data have been changed, such as access tools, devices, and architecture. But they still limited and insufficient [11] to cope with the challenges that Big data and large-scale distributed systems represent (High-performance computing as example), including scaling very quickly and flexibly.
Furthermore, with the growth of data we capture, many companies in different activity sectors (internet, e-commerce, research and science, public administrations …) have surpassed the petabytes for the storage of their data. Storing all these data become today a real challenge. Similarly, its management with the actual skills of relational databases is not suitable anymore [8].

**Data transmission:** cloud computing is generally adopted for its capacity to store a huge amount of data. Its benefit is to have an infrastructure or an applicative service that depend on the cost of what you use. This allow a interesting use flexibility (the power of storage or computing is raised according to the real need of the company) and optimize the cost by avoiding expensive investments and by releasing it also the maintenance of these infrastructures.
Due to cloud [8], the capacity of storage and the power of computing are becoming an external service. The location of the computing power and the space of storage adapted to the volume of big data allowed also its growth.

**Data analysis:** the current systems database management tools become outdated to process un-structured data that increase rapidly [11]. As Romain chaumais mention it, « classic business intelligence begin with the strategy of the company and light buttons that correspond to goals. Contrary, Big data extricate a sense in a deluge of too much data that a human cannot analyze them. These skills include analysis methods [12] called « Big analytics», used to benefit from the data

# 3. towards a new CRM and Recommendation System

## 3.1 CRM and Recommendation overview

Yore, relation between shopkeepers and customers was very strong, as they spend quality time to know each other, it's almost family relation. The shopkeeper know what his neighbor likes and sometimes needs, and neighbor is totally confident and knows that his getting good deal and no need to look elsewhere.

Mass production and mass marketing and supermarkets changed the behavior by increasing product availability, and developing new interaction way with products, however, customers lost them uniqueness, and so for the companies, they couldn't track their customers [13].

What product or service is more interesting for customers? What style and vogue prefer our customers? How shall we communicate with our client? Is a given Customer or a doubtful one?

To answer such questions, CRM (Customer Relationship Management) applications help organizations evaluate customer loyalty and profitability on measures such as repeat purchases, dollars spent, and longevity [13], In particular, customers benefit from the belief that they are saving time and money as well as receiving better information and special treatment (Kassanoff, 2000), the very important point is how could we go back to that familial relationship between companies and customers by making products tailored to fit and personalize and enhance and develop recommendation way?

Nowadays, it becomes very challenging to achieve that, for sure companies will get on because of analyzing this wealthy and fat available data, on the other hand we should take into consideration data is not the same as it was. In the first section of this paper, we presented an overview of what should we process, it is no longer about company's operational data only, it is all data, that is, all types and all sources and all size.

Today, almost all recommendation systems and CRM applications are based on structured data, represented as tables, rows and columns. Customer relationship management itself is not a new concept but is now practical due to recent advances in enterprise software technology. An outgrowth of sales force automation (SFA) tools, CRM is often referred to in the literature as one-to-one marketing (Peppers and Rogers, 1999) [13].

What we need now to enhance these systems and go to next level, is defeat paralyzed situation caused by this amount and heterogeneity of data and overcome its challenges to extract the maximum of valued key information in order to help sales forces to well put the right product on the right table.

Many related works to recommendation systems started and gave remarkable results on E-commerce; Recommender systems have as their primary task, the prediction of a value assessment of a user on a given product correspondence or buying potential and establish a list of products in a specific order.

Many methods and algorithms are used depending on what structure and type of product. In next chapters, a comprehensive and systematic study of methods and paradigms used by recommender systems, in general, recommendation algorithms use input from almost structured data about a customer's interests or trace to generate a list of recommended items Systems is carried out.

## 3.2 Content-based recommandation system

Systems implementing a content-based recommendation approach analyze a set of documents and/or descriptions of items previously rated by a user, and build a model or profile of user interests based on the features of the objects rated by that user [17].

The profile is a structured representation of user interests, adopted to recommend new interesting items. The recommendation process basically consists in matching up the attributes of the user profile against the attributes of a content object. The result is a relevance judgment that represents the user's level of interest in that object. If a profile accurately reflects user preferences, it is of tremendous advantage for the effectiveness of an information access process. For instance, it could be used to filter search results by deciding whether a user is interested in a specific Web page or not and, in the negative case, preventing it from being displayed [14].

Recent research [15] proposed that recommender systems should recommend items that maximize the users' marginal utility, instead of only items that a user likes. As the marginal utility of a camera decreases immediately after a user purchased a camera, a system's follow-up recommendation should include camera accessories instead of similar cameras. [16]

## 3.3 Collaborative filtres

The motivation of Collaborative Filtering (term introduced in 1994) is to extend the concept of word of mouth among friends to thousands of people on the Internet: friends (some people) can recommend what they liked; Internet thousands of people are likely to give you their opinion.

The objects for which you want to gauge the interest of Internet users can be of any fields: movies, restaurants, games, jokes, articles.

Collaborative filters are used to recommend cultural products essentially. And it is different, rather than concentrate on product similarity liked or bought by a given user, axe of analysis is around users and it is called users collaborative filtering, after computing users' similarity, making so a knowledge database of profiles and neighbors and take the correspondence between the given user and others and chose the strongest correlation[6] . We find also objects collaborative filtering that many algorithms applied, and models such as binary model based on the fact that user have bought/selected/checked or not a given item, evaluation model consist of grading objects. AMAZON for example used item-to-item the most popular recommendation algorithm developed by Linder *et al* [18].

If we take the following simulation:
The following table displays for each user items that purchased.

Table 1: Items purchased by users

| *Users* | *Item 1* | *Item 2* | *Item 3* |
|---|---|---|---|
| U1 | Yes | Yes | No |
| U2 | No | Yes | No |
| U3 | No | Yes | Yes |

We convert the table to a binary matrix depending on whether a purchase occurred in the past or not.

$$\begin{pmatrix} 1 & 1 & 0 \\ 0 & 1 & 0 \\ 0 & 1 & 1 \end{pmatrix}$$

Then, if a given user is interested in Item 1, for example, will suggest the item whose cosine with item 1 is the highest.

### 3.4 Combining content-based and collaborative filters

After dissection of the two approaches i.e. content-based and collaborative filters, work is carried out to derive the hybrid recommendation system, up on the strongest points of the two processes. Most hybrid recommendation systems belong to one of the following three patterns.
The first is the linear combination model, which combines the results of collaborative and content-based filters by produce two lists of prediction and combine them using adaptive weighted average.
The second is the sequential combination model, this model consist first of all of building a profiles database by means of content-based filtering based on the objects, then applying a collaborative algorithm to make expectation based on objects.
The last is the mixed combination model, in which both semantic content and ratings are applied to make recommendations these include the probabilistic model [19].

### 3.5 Sentimental Product Recommendation

In this approach many techniques interact to achieve and reach sentimental analysis from users reviews and opinions, works in [20] describe what shallow NLP techniques could provide in this concern. In general the concept of this type of recommendation is based on features. For each feature, in each sentence we extract the sentiment, by looking in a sentiment lexicon that list sentiment words. Next step is to form an opinion pattern, and then the frequency of these extracted patterns is noted. At the end we affect the sentiment that appear frequently to the features.

For each product P we now have a set of features F (P) = {f1, f2…fm} extracted from the reviews of P (Reviews (P)) and each feature Fi has an associated set of positive, negative, or neutral sentiment labels (L1, L2…). Features which are mentioned in 10% of reviews for that product are only considered and overall sentiment and popularity scores are calculated[20]; Pos(Fi, P) resp. Neg(Fi, P), Neut(Fi, P) denotes the number of positive (resp. negative, neutral) sentiment labels for feature Fi. The product case, Case (P). [21]

### 3.6 Our approach applied to the new big data paradigm

Considering the last works and the new paradigm of big data introduced in the first section, what extent big data will develop recommendation systems? Current methods and techniques will be perishable or could we re-use it?
Our work, at the top, line focuses on techniques that not depend on structured data or at least semi-structured data.
NLP techniques could provide remarkable results, however real time and volume could be an obstacle due of tools presently available; Same for text mining. But if we could chain the existing to reach that huge future of data by at least reaching the maximum numbers of the 5V's of the big data, it would be a benefit.

Solution we are focusing on it is to treat the data deluge with powerful tools and applying enhanced existing techniques to transform this data to (semi-) structured one. This work derives from data warehouse concept: as we know sources of data warehouses could be semi-structured and structured, at that juncture, we apply the ETL, to adjust this data and load it into effect-dimension tables, in brief, and it's from structured repository to another one.
Our idea is loading data from non-structured data to a semi-structured repository, that is, for example NoSQL repository, initially used to handle enormous databases for websites very large audience such as Google,

Amazon.com…, NoSQL has also spread to the bottom after 2010 Resignation functionality conventional relational DBMS in favor of simplicity. Performance remains good with scalability by simply multiplying the number of servers, reasonable solution with lower costs, especially if revenues grow along with the operation. The huge systems are the first systems affected: huge amount of data, low relational structure (or less important than the very fast access capability, even multiply servers).

Opinions may conflict with our, nonetheless the pure big-data paradigm is limited by presently tools that not support the all V's in the same tool.

## 4. Conclusions

The era of Big Data has just started, in which Data deluge is going to keep on increasing throughout the next years, and each data scientist will have to handle much more quantity of data every year. This data is becoming more diverse, larger, and faster.

Organizations must be aware of this revolution and anticipate its implementation by human and material investments. Those who will benefit from their 'Data Clients' capital, will open new perspectives towards greater competitiveness and innovation.

In this document, we discussed a brief overview on the Big data topic, including the main concerns and the main challenges for the future, as well as costumer management as an application of big data.

Big Data will allow us to extract insights that no one has extracted before. However, it is still under development and current approaches and tools are very limited to deal with the new real Big Data requirements. Further work will be focused on this corner.